\documentclass[a4paper,11pt]{article}
\usepackage{pos}
\usepackage{float}
\usepackage[utf8]{inputenc}
\usepackage{natbib}
\pdfoutput=1

\usepackage{mathtools,bbm}
\usepackage{amsfonts,amsmath}
\usepackage[percent]{overpic}
\usepackage{cancel,url,pdflscape}
\usepackage{appendix,booktabs,lscape}
\usepackage{color,xcolor,rotating}
\usepackage{tabularx, booktabs}
\usepackage{subfigure}

\newcommand{\ben}{\begin{enumerate}}
\newcommand{\een}{\end{enumerate}}
\newcommand{\bit}{\begin{itemize}}
\newcommand{\eit}{\end{itemize}}
\newcommand{\beq}{\begin{equation}}
\newcommand{\eeq}{\end{equation}}
\newcommand{\bsa}{\begin{subequations}\begin{eqnarray}}
\newcommand{\esa}{\end{eqnarray}\end{subequations}}
\newcommand{\bea}{\begin{eqnarray}}
\newcommand{\eea}{\end{eqnarray}}
\newcommand{\bean}{\begin{eqnarray*}}
\newcommand{\ean}{\end{eqnarray*}}
\newcommand{\nn}{\nonumber \\}

\newcommand{\tr}{\mbox{Tr}}
\newcommand{\fig}[1]{Fig.~\ref{#1}}
\newcommand{\eq}[1]{Eq.~(\ref{#1})}
\newcommand{\tab}[1]{Table~\ref{#1}}

\title{The static energy of a quark-antiquark pair from Laplacian eigenmodes}

\author*[a]{Roman Höllwieser}
\author[a]{Francesco Knechtli}
\author[b]{Mike Peardon}

\affiliation[a]{Dept. of Physics, Bergische Universität Wuppertal,\\
  Gaußstrasse 20, 42119 Wuppertal, Germany}

\affiliation[b]{School of Mathematics, Trinity College,\\
Dublin 2, Ireland}

\emailAdd{hoellwieser@uni-wuppertal.de}
\emailAdd{knechtli@uni-wuppertal.de}
\emailAdd{mjp@maths.tcd.ie}

\abstract{We test a method for computing the static quark-antiquark potential in lattice QCD, which is not based on Wilson loops, but where the trial states are formed by eigenvector components of the covariant lattice Laplace operator. The runtime of this method is significantly smaller than the standard Wilson loop calculation, when computing the static potential not only for on-axis, but also for many off-axis quark-antiquark separations, {\it i.e.}, when a fine spatial resolution is required. We further improve the signal by using multiple eigenvector pairs, weighted with Gaussian profile functions of the eigenvalues, providing a basis for a generalized eigenvalue problem (GEVP), as it was recently introduced to improve distillation in meson spectroscopy. We show results with the new method for the static potential with dynamical fermions and demonstrate its efficiency compared to traditional Wilson loop calculations. The method presented here can also be applied to compute hybrid or tetra-quark potentials and to static-light systems.}

\FullConference{%
 The 39th International Symposium on Lattice Field Theory, LATTICE2022
  August, 8-13, 2022
  Bonn, Germany
}

\begin{document}
\maketitle

\section{Introduction}

\begin{figure}[h]
\centering
\includegraphics[width=0.8\linewidth]{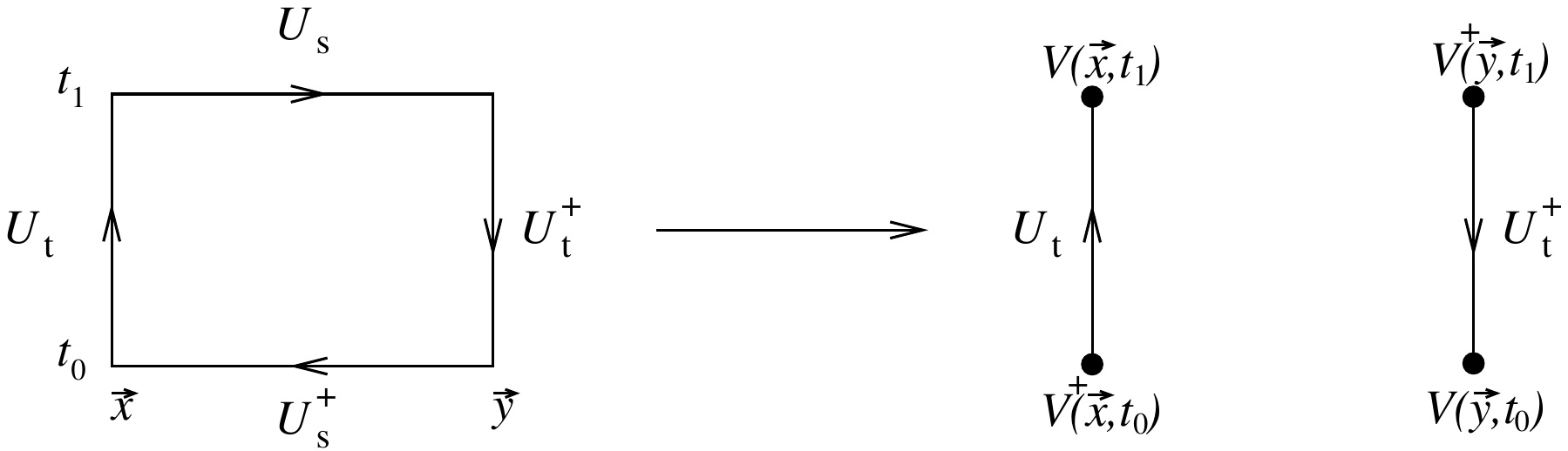}
\caption{The spatial Wilson lines $U_s(\vec x,\vec y,t)$ of the classical Wilson loop $W(R,T)$ of size $(R=|\vec y-\vec x|)\times(T=|t_1-t_0|)$ (left) can be replaced by Laplacian eigenvector pairs $V^\dagger(\vec x,t)V(\vec y,t)$ (right).}
  \label{fig:wlapl}
\end{figure}

We have developed a code for the calculation of the static potential energy based on the method presented in~\cite{Neitzel:2016lmu}, where the spatial Wilson lines $U_s(\vec x,\vec y,t)$ of a classical Wilson loop $W(R,T)$ of size $(R=|\vec y-\vec x|)\times(T=|t_1-t_0|)$ are replaced by Laplacian eigenvector pairs $V(\vec x,t)V^\dagger(\vec y,t)$\footnote{Indeed, $U_s(\vec x,\vec y,t)$ and $V(\vec x,t)V^\dagger(\vec y,t)$ have the same gauge transformation behavior.} corresponding to the lowest eigenvalues $\lambda$:
\bea
W(R,T)&=&\sum_{\vec x,t_0}\bigg\langle\tr[U_t(\vec x;t_0,t_1)U_s(\vec x,\vec y,t_1)U_t^\dagger(\vec y;t_0,t_1)U_s^\dagger(\vec x,\vec y,t_0)]\bigg\rangle\nn
&\rightarrow&\sum_{\vec x,t_0}\bigg\langle\tr[U_t(\vec x;t_0,t_1)V(\vec x,t_1)V^\dagger(\vec y,t_1)U_t^\dagger(\vec y;t_0,t_1)V(\vec y,t_0)V^\dagger(\vec x,t_0)]\bigg\rangle\,,\label{eq:neitzel}
\eea
with $U_t(\vec x,t_0,t_1)$ the temporal (static) Wilson line at space point $\vec x$ from time $t_0$ to $t_1$. Sandwiched between two eigenvectors at corresponding start- and end-times $V(\vec x,t_0)$ and $V(\vec x,t_1)$, it gives a static quark line $Q(\vec x,t_0,t_1)=V(\vec x,t_0)U_t(\vec x,t_0,t_1)V(\vec x,t_1)$ at $\vec x$ of time extent $T=|t_1-t_0|$. Its expectation value $\langle Q(\vec x,T)\rangle$ of course vanishes except for $T=0$. When combined with another static quark line $\bar Q(\vec y,t_0,t_1)$ at $\vec y$, it gives the above Wilson loop in $\eq{eq:neitzel}$ for $R=|\vec y-\vec x|$ in \fig{fig:wlapl}. 

$V(\vec x,t)V^\dagger(\vec y,t)$ represents all possible paths from $\vec x$ to $\vec y$ on the lattice, hence, we can not only form straight lines (on-axis), but also off-axis paths very easily, which would correspond to very complicated stair-like constructions of link variables. In fact, this simple method of measuring of off-axis spatial Wilson lines and loops is one of the main advantages of this method. Many off-axis separations are required for a fine resolution of the static potential which is important, {\it e.g.}, when performing a detailed investigation of string breaking \cite{Bali:2005fu,Bulava:2019iut} or when matching the perturbative and the lattice QCD static potential to determine the scale $\Lambda_{\scriptstyle \overline{\textrm{MS}}}$ \cite{Brambilla:2010pp,Jansen:2011vv,Bazavov:2012ka,Bazavov:2014soa,Karbstein:2014bsa}. 

We present an improvement of \eq{eq:neitzel} using not only the eigenvector corresponding to the lowest eigenvalue, but a number $N_v$ of lowest eigenvectors $V_i$ weighted with Gaussian profiles depending on their eigenvalues $\lambda_i$. A similar method was successfully applied to hadronic correlation functions in~\cite{Knechtli:2022bji} where an optimal smearing profile was introduced in the distillation framework~\cite{Peardon:2009gh}, which can be equivalently expressed as an optimal creation operator for a meson. In the case of the static potential we get an improvement for the static energies, which reach their plateau values at earlier temporal distances, to be quantified below. 

\section{The improved static energy based on Laplacian eigenvectors}
 
First, we write the classical Wilson loop of size $(R=|\vec x-\vec y|)\times(T=|t_1-t_0|)$ using trial states which are formed by eigenvector components of the covariant lattice Laplace operator as a transfer matrix\footnote{We thank Jeff Greensite for a fruitful discussion which lead to this slightly different approach  compared to the analysis presented in the original talk.} of $N_v\times N_v$ eigenvectors $V_i$ and $V_j$ in time slices $t_0$ and $t_1$ respectively, 
\bea
W_{ij}(R,T)&=&\sum_{\vec x,t_0}\bigg\langle\tr[V_i^\dagger(\vec x,t_0)U_t(\vec x;t_0,t_1)V_j(\vec x,t_1)V^\dagger_j(\vec y,t_1)U_t^\dagger(\vec y;t_0,t_1)V_i(\vec y,t_0)]\bigg\rangle.\label{eq:wij}
\eea
We could either take a double sum over all eigenvector pairs $i,j=1\ldots N_v$, which increases the statistics and the signal of the Wilson loops or we can solve a GEVP for the Wilson loop basis matrix $W_{ij}$, which however is very ill-conditioned. We therefore prune $W_{ij}$ using the three most significant singular vectors $u_k$ from a singular value decomposition\footnote{$W_{ij}=UDV^\dagger$ with $U$ and $V$ being unitary matrices, whose column vectors $u_k$ and $v_l$ form an orthonormal basis, and $D$ being diagonal with non-negative real numbers on the diagonal.} via $\tilde W_{kl}=u_k^\dagger W_{ij}u_l$, which keeps only (a combination) of useful operators and improves the stability of the GEVP for fixed $R/a$: $\tilde W(t)v_k(t,t_0)=\rho_k(t,t_0)\tilde W(t_0)v_k(t,t_0)$. From the principal correlators $\rho_k(t,t_0)$ we get the effective energies/masses. From the vectors $v_k(t,t_0)$ and $u_k$ we see that the GEVP favors low-lying eigenmodes. On the other hand, an increasing number $N_v$ of eigenvectors enhances the signal and in particular the overlap with the ground-state in the effective energies/masses for small distances. 

\begin{figure}[h]
\centering
\includegraphics[width=0.495\textwidth]{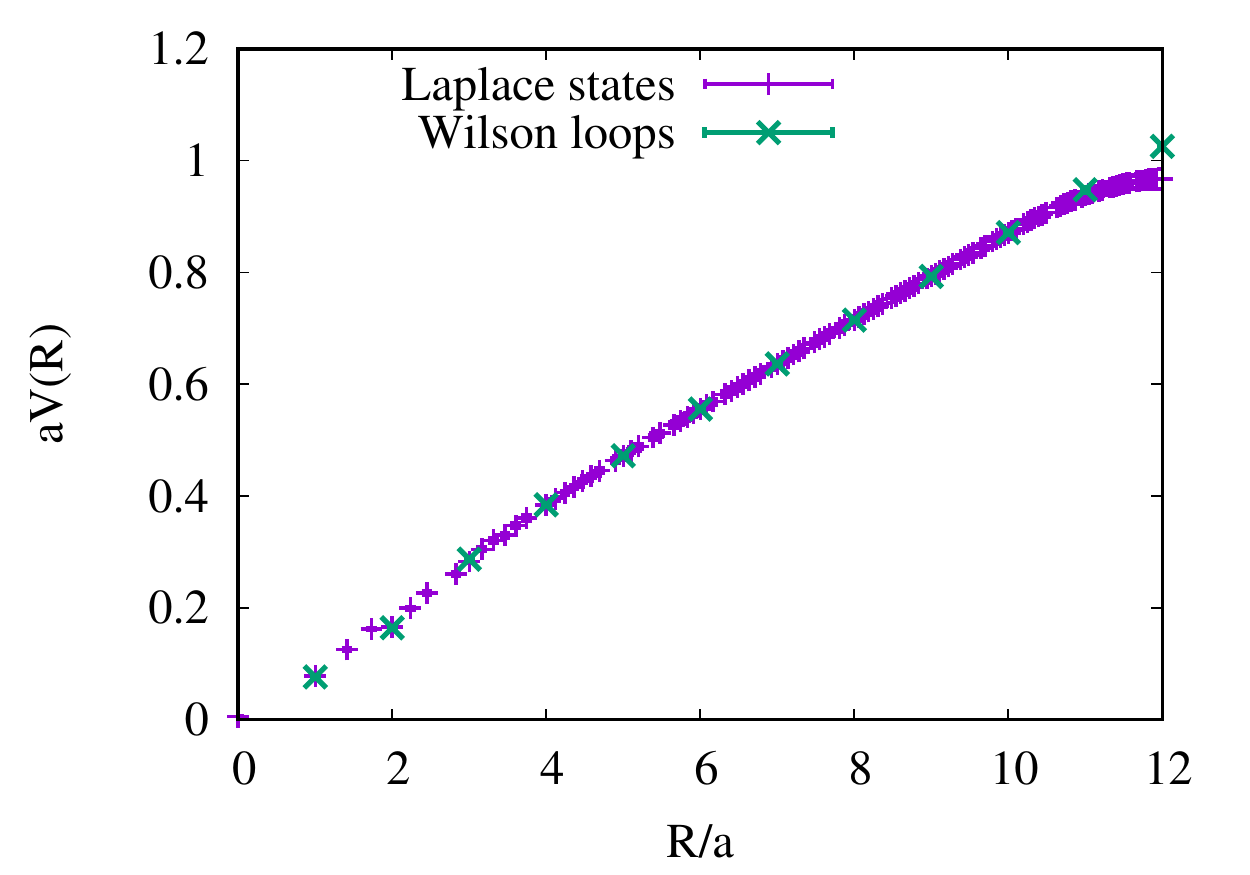}
\includegraphics[width=0.495\linewidth]{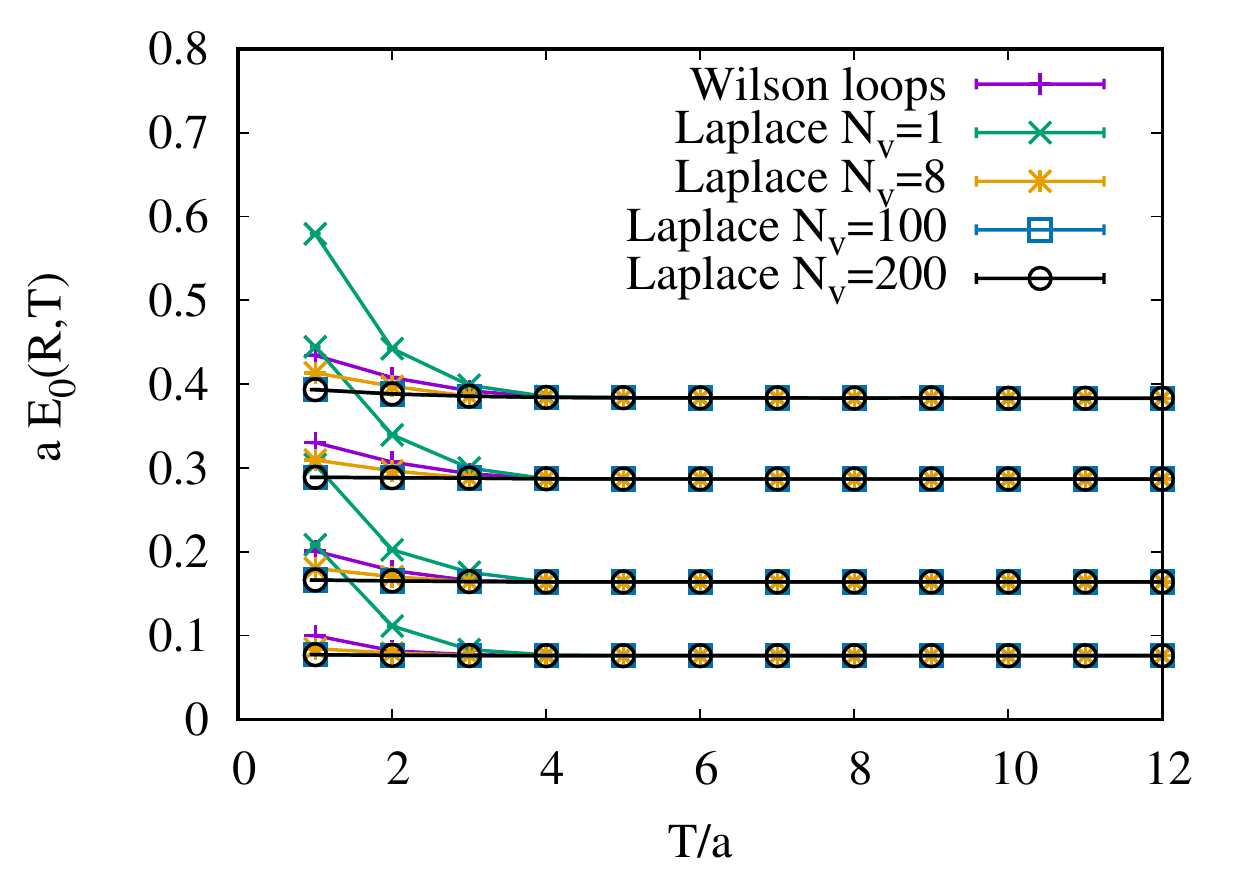}
\caption{The static potential (left) and the first four effective energies/masses (right) on ensemble Em1 using the Laplacian eigenvector approach. The first is computed for all (!) on- {\bf and off}-axis separations $R/a$ and $N_v=8$, showing a flattening at half the lattice size $R/a=12$ (left). The ground state overlap can be drastically improved by using more eigenvectors, see \tab{tab:ovl}, we get earlier plateaus for larger $N_v$.}
  \label{fig:vwlEm1}
\end{figure}

In \fig{fig:vwlEm1} we show our results for the new observable compared to actual Wilson loop measurements on a $24^3\times48$ lattice ensemble at $\beta=5.3$ ($a=0.0658$fm) and $N_f=2$ non-perturbatively $O(a)$-improved Wilson quarks with $\kappa=0.13270$, corresponding to half the charm quark mass. The original Wilson loop was measured on 4646 gauge configurations, while $W_{ij}$ was measured on every fourth configuration only (1160 measurements). In the left plot we present the static potential $V(R,T)=\lim_{T\rightarrow\infty}\log[W(R,T)/W(R,T+1)]$ for all on- and off-axis separations $R/a$ from $N_v=8$ Laplacian eigenvector pairs compared to on-axis Wilson loop results. In the measurement of 
Wilson loops all gauge links are HYP2 smeared~\cite{Donnellan:2010mx}. We observe a discrepancy of the two methods only for large $R/a$. In fact, at $R=12a$ (half the lattice size) the force between $Q\bar Q$ must vanish due to symmetry, {\it i.e.}, the static potential must be flat, in agreement with the new numerical results.
This effect however goes away for increasing $N_v$ and we get the exact same potential as for Wilson loops. The right plot in~\fig{fig:vwlEm1} clearly shows that an increasing number $N_v$ of Laplacian eigenvector pairs improves the ground state overlap of the effective energies/masses for the static quark-anti-quark system. Already $N_v=8$ eigenvector pairs reach the plateau values faster than the original Wilson loops, while at $N_v=100$ the effect seems to saturate, we do not see a difference between $N_v=100$ and $N_v=200$. The ground state overlaps can be quantified by taking the $t$-average over the mass-plateau region of 
\bea
\frac{W(R,t)}{W(R,t_0)}\frac{\cosh \left( \left( \frac{T}{2} - t_0 \right) am_0(R) \right)}{\cosh \left( \left( \frac{T}{2} - t \right) am_0(R) \right)},\label{eq:ovl}
\eea
using the same $t_0=3$ as in the GEVP and corresponding ground state masses $am_0(R)$ from a $\cosh$-fit. These so-called 'fractional overlaps' are listed in \tab{tab:ovl} and underpin again, that a large number $N_v$ of eigenvector pairs gives better overlaps for small distances $R/a$, but with decreasing importance especially for large distances, where already $N_v<100$ shows better overlaps. 

\begin{table}[h]
\centering
\begin{tabular}{c|cccccc|cc}
$R/a$ & $N_v=1$ & $8$ & $64$ & $100$ & $200$ & Gauss & Wloop & GEVP \\
\midrule
1 & 0.773(3) & 0.945(1) & 0.970(1) & 0.980(1) & 0.982(1) & 0.993(1) & 0.921(1) & 0.983(1) \\
2 & 0.747(4) & 0.929(2) & 0.964(1) & 0.988(1) & 0.987(1) & 0.989(1) & 0.891(1) & 0.978(1) \\
3 & 0.723(4) & 0.878(2) & 0.984(2) & 0.987(2) & 0.986(1) & 0.988(1) & 0.867(1) & 0.972(2) \\
4 & 0.726(5) & 0.874(3) & 0.921(2) & 0.982(2) & 0.984(2) & 0.986(2) & 0.841(2) & 0.965(3) \\
5 & 0.637(6) & 0.871(4) & 0.979(3) & 0.983(3) & 0.982(3) & 0.983(3) & 0.813(2) & 0.956(5) \\
6 & 0.629(6) & 0.869(4) & 0.978(4) & 0.981(4) & 0.980(3) & 0.981(3) & 0.793(3) & 0.948(6) \\
7 & 0.619(7) & 0.869(5) & 0.977(4) & 0.982(4) & 0.979(4) & 0.987(4) & 0.772(3) & 0.934(7) \\
8 & 0.598(8) & 0.862(6) & 0.972(5) & 0.971(5) & 0.970(4) & 0.974(4) & 0.745(4) & 0.953(8) \\
9 & 0.572(8) & 0.857(6) & 0.960(5) & 0.954(5) & 0.934(4) & 0.963(3) & 0.708(4) & 0.947(9) \\
10 & 0.540(9) & 0.840(7) & 0.955(5) & 0.941(6) & 0.931(5) & 0.965(1) & 0.671(5) & 0.94(1) \\
11 & 0.426(9) & 0.807(7) & 0.943(6) & 0.934(5) & 0.93(1) & 0.956(9) & 0.649(4) & 0.93(1) \\
12 & 0.33(7) & 0.79(2) & 0.94(1) & 0.932(9) & 0.92(1) & 0.95(1) & 0.64(2) & 0.92(1)
 \end{tabular}
 \caption{Fractional overlaps with the corresponding ground state masses $am_0(R)$. In general, an increasing number $N_v$ of Laplacian eigenvectors enhances the overlap up to about $N_v\approx100$ and are already better for $N_v=8$ than standard Wilson loops (column 7). The overlaps for Laplacian trial states from a GEVP with Gaussian profiles in the 6th column are better than Wilson loop results from a GEVP with different spatial HYP smearing levels (column 8), see text below.}\label{tab:ovl}
 \end{table}

\begin{figure}[h]
\centering
\includegraphics[width=0.522\textwidth]{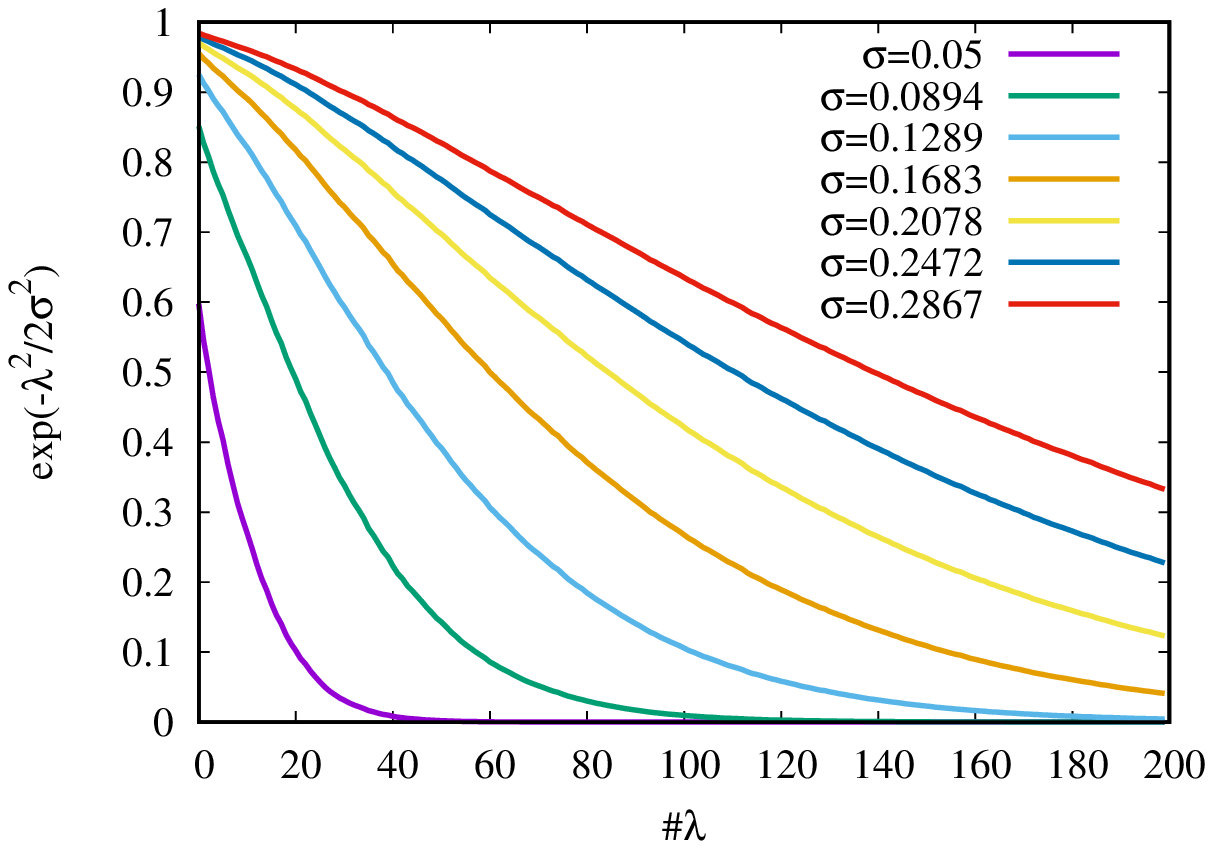}
\includegraphics[width=0.472\linewidth]{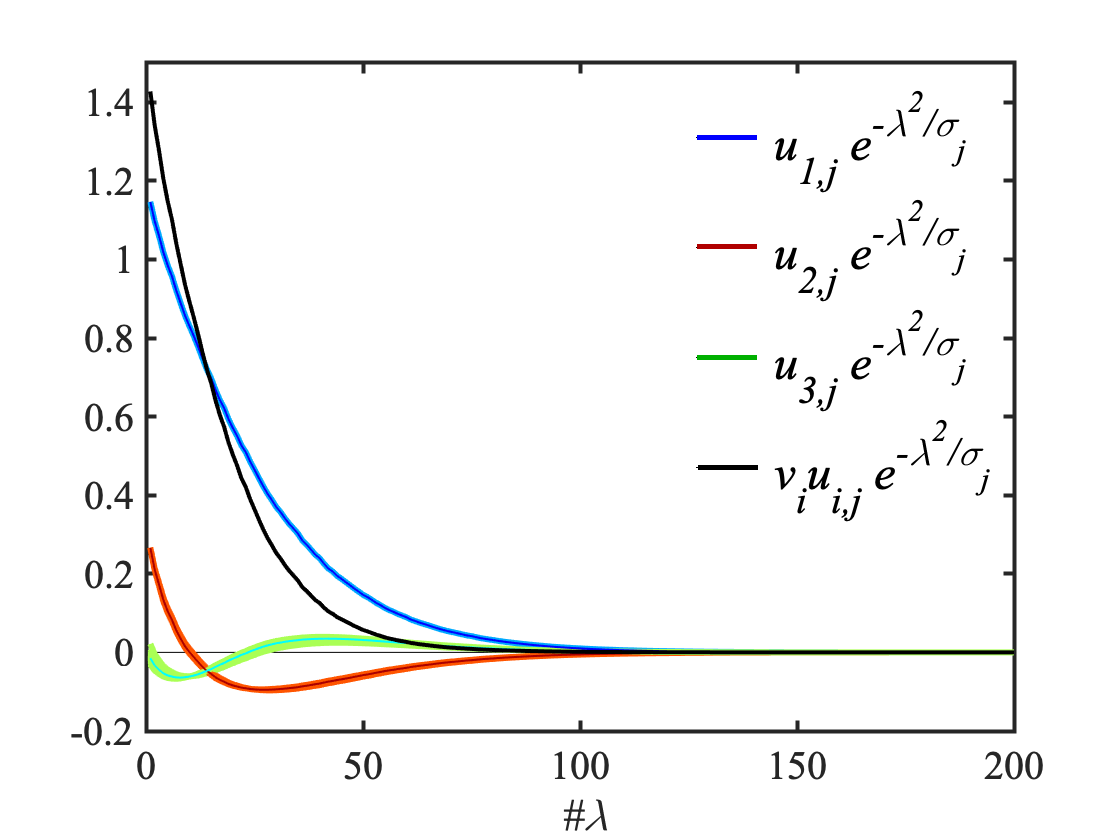}
\caption{The Gaussian profiles (left) pruned by SVD vectors to 'optimal' profiles (right) for the GEVP.}
  \label{fig:propti}
\end{figure}

Therefore, instead of feeding a large ill-conditioned $N_v\times N_v$ transfer matrix to the GEVP we introduce Gaussian profile functions $\exp(-\lambda^2/2\sigma^2)$ for each eigenvector with the corresponding eigenvalue $\lambda$ and different Gaussian widths $\sigma$. We define the new GEVP basis matrix
\bea
W_{kl}(R,T)&=&\sum_{i,j}^{N_v}N_{kl}(\lambda_i,\lambda_j)W_{ij}(R,T),\label{eq:wkl}
\eea
by (double-)summing over the $N_v$ eigenvector pairs\footnote{Actually, the sum over $t_0$ in \eq{eq:wij} should be the outer most in \eq{eq:wkl}, since the eigenvalues minimally vary between time-slices, however using the average eigenvalues over all time-slices does not change the final results or precision.}, weighted by Gaussian profile functions \[N_{kl}(\lambda_i,\lambda_j)=\exp(-\lambda_i^2/2\sigma_k^2)\exp(-\lambda_j^2/2\sigma_l^2)\] with $\sigma_{k,l}\in[0.05,0.0894,0.1289,0.1683,0.2078,0.2472,0.2867]$, see \fig{fig:propti} (left). This way we gain statistics (precision) by the double sum and find an optimal number of 'important' eigenvectors by solving the GEVP for the Wilson loop basis matrix $W_{kl}$ using combinations of profiles with different $\sigma_{k,l}$. Again, we first prune $W_{kl}$ using the three most significant singular vectors $u_m$ via $\tilde W_{mn}=u_m^\dagger W_{kl}u_n$, which keeps only (a combination) of useful operators and improves the stability of the GEVP for fixed $R/a$: $\tilde W(t)v_i(t,t_0)=\rho_i(t,t_0)\tilde W(t_0)v_i(t,t_0)$. From the principal correlators $\rho_i(t,t_0)$ we get the effective energies/masses. From the singular vectors $u_i$ we get the pruned or 'optimal' profiles $\sum_ju_{i,j}\exp(-\lambda^2/\sigma_j)$, depicted in blue, red and green in \fig{fig:propti} (right) for $R/a=3$, together with the 'optimal' ground state profile, the linear combination of pruned profiles using the generalized eigenvectors $v_i$, $\sum_{i,j}v_iu_{i,j}\exp(-\lambda^2/\sigma_j)$ in black. Indeed, the 'optimal' profiles give us a number $N_v\approx100$ of 'important' eigenvectors for $R/a=3$, for larger distances this number slightly decreases.

\section{Results \& Timing}

\begin{figure}[h]
\centering
\includegraphics[width=0.51\textwidth]{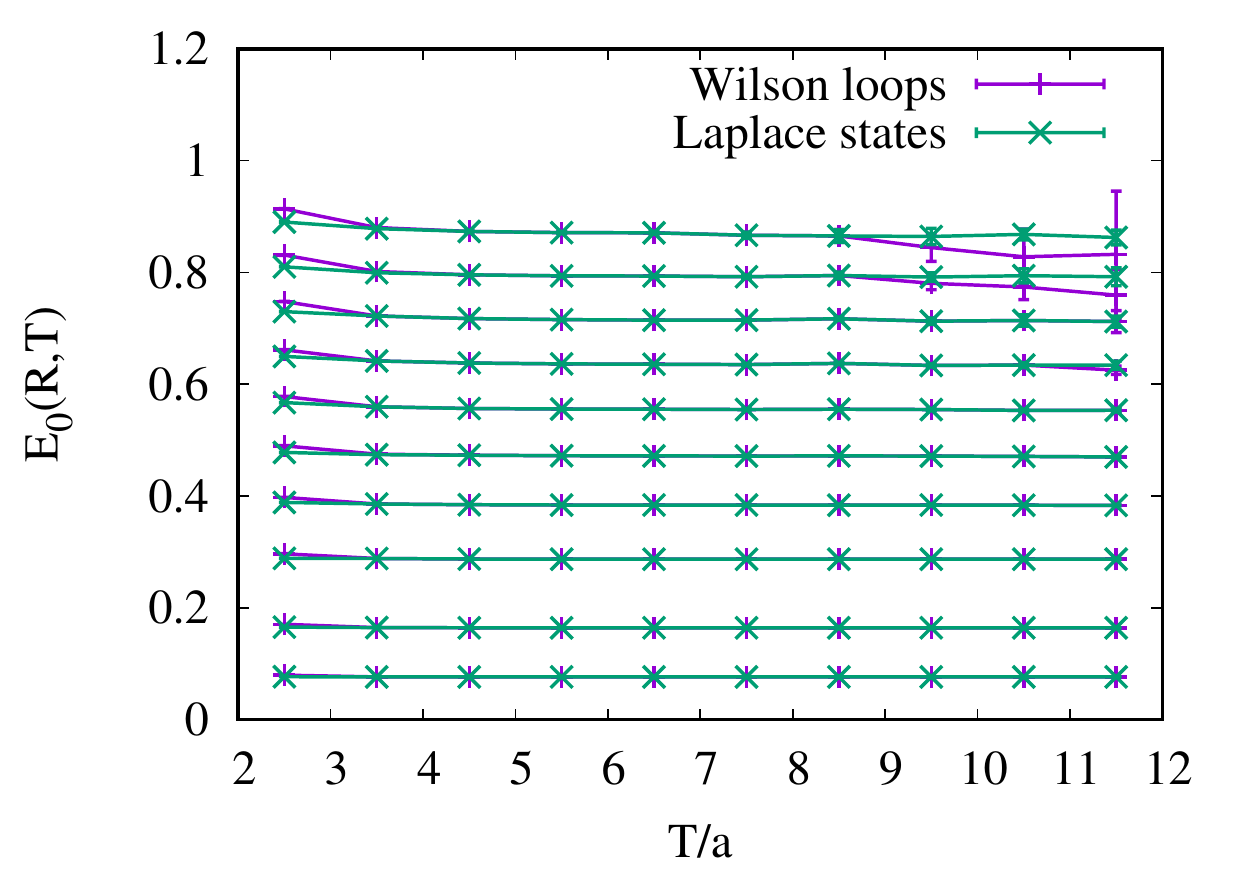}
\includegraphics[width=0.484\textwidth]{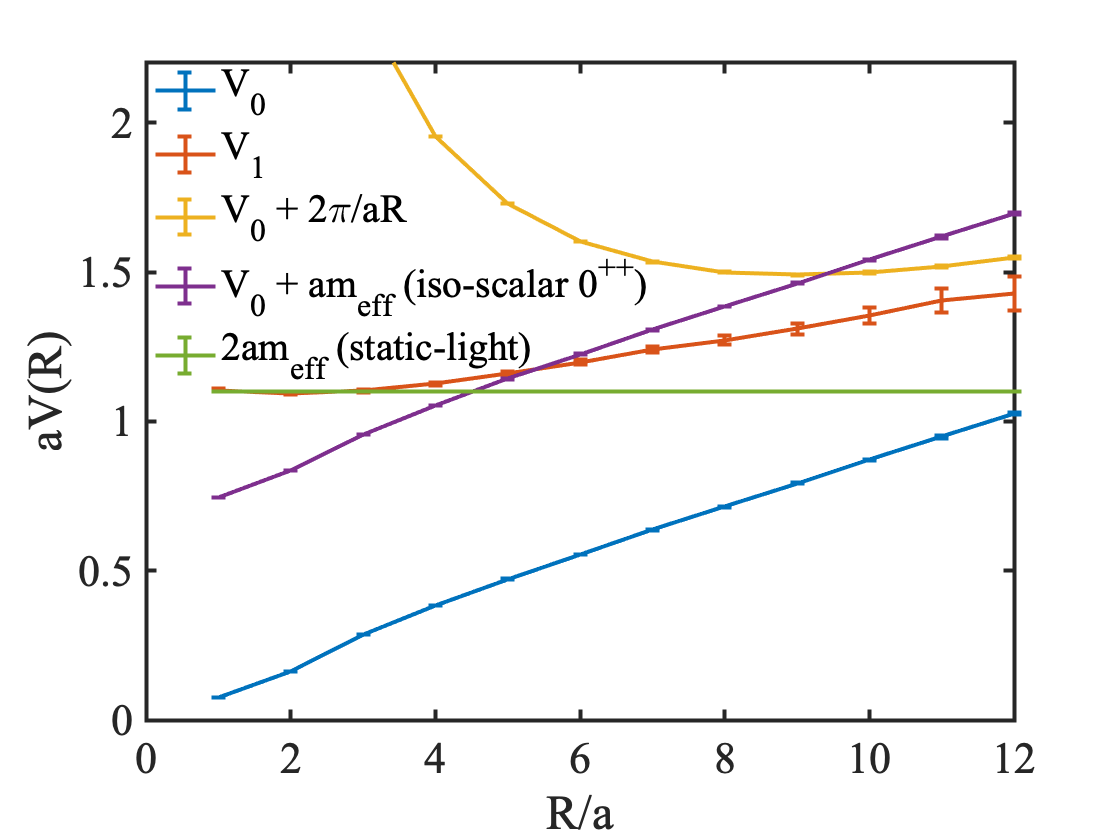}
\caption{The effective energies/masses (left) using the Laplacian eigenvector approach with Gaussian profiles and Wilson loops with spatial HYP smearing after solving the GEVP, and static potentials $V_n$ (right) for the ground ($n=0$) and first excited ($n=1$) states, which we compare with the excited string state $V_0+2\pi/aR$, the lowest $0^{++}$ iso-scalar meson (possible glueball) state $V_0+am_{\rm eff}(\text{iso-scalar }0^{++})$ from~\cite{Knechtli:2022bji} and two times the static-charm meson mass on the $N_f=2$ ensemble.}
  \label{fig:res}
\end{figure}

With the method presented in the previous section, the implementation of~\cite{Neitzel:2016lmu}, see also \eq{eq:neitzel}, using only the eigenvector corresponding to the lowest eigenvalue, can be significantly improved by (double-)summing over the lowest $N_v$ eigenvector pairs, weighted by Gaussian profile functions using their corresponding eigenvalues $\exp(-\lambda^2/2\sigma^2)$ and different Gaussian widths $\sigma_{k/l}$. Just like for the standard Wilson loop, where we solve a generalized eigenvalue problem for the correlation matrix of Wilson loops with different spatial smearing levels, we feed $W_{kl}$ from \eq{eq:wkl} (or its pruned version) into a GEVP which gives us the 'optimal' profiles or most important eigenvector pairs for each $R/a$. We present the improvement of effective energies using our method in the left plot of \fig{fig:res}, showing the effective energies/masses using the improved Laplacian eigenvector approach with Gaussian profiles after solving the GEVP together with smeared Wilson loop results. In fact, the results from Laplacian modes show higher accuracy than those from Wilson loops, even though measured only on a fourth of the total statistics, however we increased the averaging by the double-sum over different eigenvectors. 

The right plot of \fig{fig:res} presents the static potentials $V_n$ for the ground ($n=0$) and first excited ($n=1$) states using the Laplacian eigenvector approach with Gaussian profiles after solving the GEVP. The first excited state ($n=1$) is just included to show the potential of the method, we compare it with the excited string state $V_0+2\pi/aR$, the lowest $0^{++}$ iso-scalar meson (possible glueball) state $V_0+am_{\rm eff}(\text{iso-scalar }0^{++})$ from~\cite{Knechtli:2022bji} and two times the static-charm meson mass which were also evaluated using the new method in combination with 'charm-perambulators' also from~\cite{Knechtli:2022bji} on the same $N_f=2$ ensemble.

The computational effort of this new method is even favorable to the standard Wilson loop calculation, especially for off-axis separations. In fact, for our test ensemble on a $24^3\times48$ lattice the computation of on-axis Wilson loops using 4 spatial smearing levels (0, 10, 20, 30 HYP steps)~\cite{Donnellan:2010mx} is equally expensive as the calculation of 100 Laplacian eigenvectors and Laplace states with 3 Gaussian profiles including off-axis distances!

\section{Conclusions \& Outlook}

We presented an alternative operator for a static quark-anti-quark pair based on Laplacian eigenmodes and improved the operator given in~\cite{Neitzel:2016lmu} using a large number of eigenvectors weighted with Gaussian profiles. We observe earlier plateaus in the effective masses and a better signal. The main advantage of this eigenvector approach however is to have an efficient method to compute the static potential not only for on-axis, but also for many off-axis quark-antiquark separations. Using the standard gauge link approach for the computation of Wilson loops, is rather time consuming, since a large number of stair-like gluonic connections has to be computed (cf.\ {\it e.g.}\ \cite{Bali:2005fu} for a discussion of how to compute such off-axis Wilson loops). In comparison, computing many off-axis separations of the static potential using Laplacian eigenvectors requires less computing time, since the eigenvector components of the covariant lattice Laplace operator have to be computed only once and can then be used for arbitrary on-axis and off-axis separations without the need to compute stair-like connections. 

We want to adapt the method to also measure hybrid static potentials relevant for exotic mesons, where the gluonic string excitations (gluonic handles in the standard Wilson loop approach) can be realized by covariant derivatives of the Laplacian eigenvectors in \eq{eq:neitzel}. Finally, when we combine our static quark line with a perambulator from~\cite{Knechtli:2022bji}, we can build a static-light quark meson. The long-term plan is to put together all building blocks for observation of string breaking in QCD (mixing matrix of static and light quark propagators) in the framework of distillation~\cite{Peardon:2009gh}.

\section*{Acknowledgements} The authors gratefully acknowledge the Gauss Centre for Supercomputing e.V. (www.gauss-centre.eu) for funding this project by providing computing time on the GCS Supercomputer SuperMUC-NG at Leibniz Supercomputing Centre (www.lrz.de). The work is supported by the German Research Fund (DFG) research unit FOR5269 "Future methods for studying confined gluons in QCD" and the NRW-FAIR network, funded by the Ministry for Culture and Science of the State of North Rhine-Westphalia (MKW). For valuable discussions we thank Jeff Greensite, Tomasz Korzec and especially Juan Andrés Urrea-Niño, who also provided the Laplacian eigenvectors used in this work.

\bibliographystyle{utphys} 
\bibliography{PoS}

\end{document}